\documentclass[twocolumn,showpacs,preprintnumbers,footinbib,prd,superscriptaddress,groupedaddress,10pt]{revtex4-1}

\usepackage[utf8]{inputenc}

\makeatletter
\def\l@subsubsection#1#2{}
\def\l@subsubsubsection#1#2{}
\makeatother

\usepackage{graphicx,amssymb,amsmath,amsthm,amsfonts}
\usepackage[usenames]{color}

\usepackage{notes2bib}

\usepackage{aas_macros}
\usepackage{bm}
\usepackage{dcolumn}
\usepackage{latexsym}
\usepackage{rotating}
\usepackage{longtable}

\usepackage{enumerate}
\usepackage{tensor,multirow}
\usepackage{url}
\usepackage[linktocpage]{hyperref}

\def\be{\begin{equation}}
\def\ee{\end{equation}}
\def\beq{\begin{eqnarray}}
\def\eeq{\end{eqnarray}}
\begin{document}

\title{Collective scalarization or tachyonization: when averaging fails}

\author{Vitor Cardoso}
\affiliation{CENTRA, Departamento de F\'{\i}sica, Instituto Superior T\'ecnico -- IST, Universidade de Lisboa -- UL,
Avenida Rovisco Pais 1, 1049 Lisboa, Portugal}
\affiliation{Waseda Institute for Advanced Study (WIAS), Waseda University, Shinjuku, Tokyo 169-8050,  Japan}
\author{Arianna Foschi}
\affiliation{CENTRA, Departamento de F\'{\i}sica, Instituto Superior T\'ecnico -- IST, Universidade de Lisboa -- UL,
Avenida Rovisco Pais 1, 1049 Lisboa, Portugal}
\author{Miguel Zilh\~ao}
\affiliation{CENTRA, Departamento de F\'{\i}sica, Instituto Superior T\'ecnico -- IST, Universidade de Lisboa -- UL,
Avenida Rovisco Pais 1, 1049 Lisboa, Portugal}

\begin{abstract} 
Certain scalar-tensor theories of gravity provide negative-energy, tachyonic modes to a fundamental scalar inside matter, giving rise to non-perturbative phenomena around compact stars. Studies of this and other tachyonic instabilities always average over local matter properties. We use elementary, flat space models to understand possible collective effects and the accuracy of the averaging procedure.  
In particular, we consider bodies made of elementary constituents
which do not, in isolation, scalarize because their compactness ${\cal C}$ is too small, ${\cal C}\lesssim {\cal C}_{\rm crit}$. We show that when the individual constituents have compactness smaller but close to the threshold, one is able to scalarize composite bodies through collective effects, and the compactness of the composite body can be made arbitrarily small.
On the other hand, our results suggest that when the fundamental building blocks have very low compactness, then scalarization
of the composite body requires a global compactness ${\cal C}_{\rm global}\gtrsim {\cal C}_{\rm crit}$. Thus, our results rule out scalarization of dilute bodies via collective effects.
\end{abstract}

%%%%
%%%%
\maketitle

%%%%%%%%%%%%%%%%%%%%%%%%%%%%%%
%\section{Introduction}
%%%%%%%%%%%%%%%%%%%%%%%%%%%%%%
%%%%%%%%%%%%%%%%%%%%%%%%%%%%%%%%%%%%%%
\noindent{\bf{\em Introduction.}}
%\section{Introduction}
%%%%%%%%%%%%%%%%%%%%%%%%%%%%%%%%%%%%%%
Our understanding of the universe is incomplete, even at coarse-grained level. 
A substantial fraction of the required energy budget necessary to explain all the astrophysical and cosmological observations
seems to be of unknown nature. If the ``Standard Model'' of particle physics -- along with its numerous species content -- is a good guide, then one may expect a plethora of new fundamental fields to be a part of the dark sector of our universe~\cite{Arvanitaki:2009fg,Marsh:2015xka}. 
The reasons for their elusiveness may rest with their very weak coupling to standard model particles; in such a case, very precise and sensitive measurements are necessary to probe their existence; alternatively, the equivalence principle suggests universal behavior even in strong gravity, where new effects such as superradiant growth of structures around spinning black holes or stars are possible~\cite{Brito:2015oca}.

Baryonic matter could control the behavior or ``awakening'' of such new degrees of freedom in different ways. An appealing 
framework concerns scalar-tensor theories, where the new degree of freedom is a non-minimally coupled scalar field $\Phi$. In a wide class of these theories, $\Phi$ is governed by a modified Klein-Gordon (KG) equation of the form~\cite{Damour:1992we,Damour:1993hw}
%
%\be
$\nabla_\mu\nabla^\mu\Phi=\alpha T \Phi$,
%\ee
%
after a proper fluctuation around some background is performed. Here $T$ is the trace of the local matter stress-energy tensor.
The essence of the mechanism as well as the practical issues we wish to discuss are all present in flat space and dilute configurations. These issues concern a negative coupling constant $\alpha$. Thus, we wish to focus here on the evolution equation for a scalar field in flat space, when it acquires an effective negative mass-squared $\mu_{\rm eff}^2=-\mu^2$ in regions of nontrivial matter content with density $\rho$,
\be
\nabla_\mu\nabla^\mu\Phi=\alpha \rho \Phi\equiv -\mu^2\Phi\,.
\ee
Our discussion and results will therefore carry over to any theory with (position-dependent) tachyonic modes.

%%%%%%%%%%%%%%%%%%%%%%%%%%%%%%%%%%%%%%%%%%%%%%%%%%%%%%%%%%%%%%
%\subsection{Tachyons in matter and spontaneous scalarization}
%%%%%%%%%%%%%%%%%%%%%%%%%%%%%%%%%%%%%%%%%%%%%%%%%%%%%%%%%%%%%%
%%%%%%%%%%%%%%%%%%%%%%%%%%%%%%%%%%%%%%%%%%%%%%%%%%%%%%%%%%%%%%%%%%%%%
\noindent{\bf{\em Scalarization and tachyons in matter.}}
%%%%%%%%%%%%%%%%%%%%%%%%%%%%%%%%%%%%%%%%%%%%%%%%%%%%%%%%%%%%%%%%%%%%%
The above theory admits a trivial scalar $\Phi=0$ as a solution, but such solution may be unstable. One of the simplest toy models consists on a spherically symmetric matter distribution of radius $R$of constant density $\rho_0$, modeling for example a planet or star. In spherical coordinates and spherical symmetry, with $\Phi=e^{-i\omega t}\Psi(r)$, the KG equation takes the form
\be
\frac{1}{r}\frac{d^2}{dr^2}(r\Psi)+k^2\Psi=0\,,
\ee
where $k^2=\omega^2+\mu^2$. The regular solution at the center is $\Psi=A\sin(kr)/r$. Outside the object, one looks for outgoing regular modes, $\Psi=Be^{i\omega r}/r$. Matching the field and its derivative, one finds the characteristic equation
%
%\be
$\sin kR=\pm k /\mu$.
%\ee
%
It is easy to show that an instability is present iff~\cite{Landau:book_fieldtheory}~\footnote{This is a special case of a more general result: the polarizability of a sphere of radius $R$ in this theory, defined as the ratio between the external and induced $\ell-$ pole, is $R^{2\ell+1}J_{\ell+3/2}(R\mu)/J_{\ell-1/2}(R\mu)$ with $J(x)$ a Bessel function. For $\ell=0$ there's a pole at $R\mu=\pi/2$ signaling a new regular solution.}
\be
\mu^2R^2>\frac{\pi^2}{4}\,.\label{eq_instability_sphere}
\ee
Using the identity $\mu^2=-\alpha\rho_0$, this result can be expressed in terms of the compactness ${\cal C}$ of the sphere,
\be
{\cal C}\equiv \frac{M}{R}>-\frac{\pi^3}{3\alpha}\equiv {\cal C}_{\rm crit}\,,
\ee
with $M$ the total mass of the sphere.

Two aspects are noteworthy. For large $R$ and fixed density, an instability sets in, and the final state is in general a solution with a non-trivial scalar profile. We term this process ``scalarization,'' since the trivial solution with a vanishing scalar is not stable. The understanding of the final state is outside the scope of this work, and requires knowledge about how the scalar backreacts on the matter distribution. Such stationary solutions were extensively studied within the context of non-minimally coupled scalars or vectors~\cite{Damour:1992we,Damour:1993hw,Harada:1997mr,Lima:2010na,Pani:2010vc,Cardoso:2013fwa,Cardoso:2013opa,Ramazanoglu:2016kul,Ramazanoglu:2017xbl,Annulli:2019fzq}. Second, the mechanism can be made to work for black holes as well, if instead of couplings to matter density one considers higher-curvature gravitational corrections~\cite{Doneva:2017bvd,Silva:2017uqg,Antoniou:2017acq} or axionic-type potentials~\cite{Boskovic:2018lkj}. All that is going to be discussed here applies qualitatively but not quantitatively to such setups as well.
Most importantly of course, is the existence of a threshold radius below which $\Phi=0$ is the only and stable solution.
It teaches us a lesson about tachyons: by itself a negative but localized mass-squared is not synonymous of an instability.
From Eq.~\eqref{eq_instability_sphere} we see that meaningful and interesting constraints on small $|\alpha|$ require very compact matter configurations.

%%%%%%%%%%%%%%%%%%%%%%%%%%%%%%%%%%%%%%%%%%%%%%%%%%%%%%%%%%%
%\subsection{Issues with our understaning of scalarization}
%%%%%%%%%%%%%%%%%%%%%%%%%%%%%%%%%%%%%%%%%%%%%%%%%%%%%%%%%%%
%%%%%%%%%%%%%%%%%%%%%%%%%%%%%%%%%%%%%%%%%%%%%%%%%%%%%%%%%%%%%%%%%%%%%
\noindent{\bf{\em Issues with our understanding of scalarization.}}
%%%%%%%%%%%%%%%%%%%%%%%%%%%%%%%%%%%%%%%%%%%%%%%%%%%%%%%%%%%%%%%%%%%%%
The above conclusions can be called into question, and should be clearly understood for different reasons:

\paragraph{The universality of constraint~\eqref{eq_instability_sphere}.} Our interest was prompted by a simple setup, that of an infinite slab of matter of thickness $L$. This essentially one-dimensional problem, together with the requirement of regularity, yields characteristic modes as roots of the equation
\beq
2i\omega \cos{(L\sqrt{\omega^2+\mu^2})}=-\frac{(2\omega^2+\mu^2)\sin{(L\sqrt{\omega^2+\mu^2})}}{\sqrt{\omega^2+\mu^2}}\,.\nonumber
\eeq
It is easy to see that at small $L\mu,\omega/\mu$ one can expand the equation above and get the solution 
\be
\omega=i\frac{L}{2}\mu^2\,.\label{slab}
\ee
Thus, an instability is always present, in line with the result by Ref.~\cite{Buell}, 
since the effective potential $V=-\mu^2<0$ here. 
Note that this does not contradict Eq.~\eqref{eq_instability_sphere} which refers instead to a three-dimensional potential, and therefore boundary conditions at $r=0$ require finiteness of the field~\footnote{In essence, Ref.~\cite{Buell} shows that the {\it one-dimensional} Schrodinger equation with a negative-definite potential must have bound states; it is easy to show that this implies that the corresponding Klein-Gordon equation has instabilities, since it is described by a potential $V=-\mu^2$. Incidentally, this is also the reason why placing a black hole in the system may sometimes favour scalarization, as was serendipitously observed in Ref.~\cite{Cardoso:2013fwa}.}. However, it does call into question the generality of any result with a threshold radius. In other words, there is some tension between the result in Eq.~\eqref{eq_instability_sphere} and that of Eq.~\eqref{slab}. 
These results can be made compatible of course, if one realizes that the geometry of the system is important, which leads one to the next issue.

\paragraph{The averaging procedure for composite bodies.} Composite bodies have a fine structure, dictated by their composition. For example, planets and stars are microscopically described
by a density (and therefore $\mu^2$) which varies periodically, large close to atomic nuclei and negligibly small away from these. Such fine structure is impossible to model in current studies, due to the large number of particles involved. Thus, it is approximated by a local average of the star properties. What is the impact of averaging on the scalarization properties of a body? In particular, does Eq.~\eqref{eq_instability_sphere} also hold for bodies with structure?
Can such bound be violated when a large number of constituents is considered?

\paragraph{Dynamical scalarization of neutron stars.} That geometrical effects can be important is suggested by previous studies. It was shown, via dynamical simulations of a binary inspiral in General Relativity, that the presence of a second body may facilitate scalarization~\cite{Palenzuela:2013hsa}. This is a hint towards new ``collective'' phenomena when a large number of bodies is present.

\paragraph{Solid-state physics and periodic lattices.} Finally, there is an interesting connection between investigating instabilities in the presence of a large number of objects and investigating bound states (in Schrodinger's equation) of large lattices of ions. The latter has a long history, with some well known results in one-dimension, such as the Kronig-Penney results for band gaps~\cite{Kronig:1931,PhysRevB.33.2122}. Although an analysis of new, lattice-triggered bound states is not usually reported, there are examples of new ``global'' bound states not allowed by ions~\cite{Demkov:1970}. This raises the interesting prospect that a large number of atoms, say in a chair or in a planet, might permit scalarization when a naive analysis based on its average properties would predict otherwise.

%%%%%%%%%%%%%%%%%%%%%%%%%%%%%%%%%%%%%%%%%%%%%%%%%%%%%%%%%%%%%%%%%
%\section{Three different toy models for collective scalarization}
%%%%%%%%%%%%%%%%%%%%%%%%%%%%%%%%%%%%%%%%%%%%%%%%%%%%%%%%%%%%%%%%%
%%%%%%%%%%%%%%%%%%%%%%%%%%%%%%%%%%%%%%%%%%%%%%%%%%%%%%%%%%%%%%%%%%%%%
\noindent{\bf{\em Collective scalarization.}}
%%%%%%%%%%%%%%%%%%%%%%%%%%%%%%%%%%%%%%%%%%%%%%%%%%%%%%%%%%%%%%%%%%%%%
To summarize, Eq.~\eqref{eq_instability_sphere} suggests that astrophysical objects can only scalarize for $|\alpha| \gtrsim 1$  if they are neutron stars. Our plan here is to study composite configurations, and to understand if their ``scalarization properties'' are well described by an averaging procedure. In other words, if condition \eqref{eq_instability_sphere} is correct on an average sense, then for a composite body of total mass $M_{\rm tot}$ and size ${\cal L}$, scalarization should be tied to the condition $-\alpha \rho {\cal L}^2\gtrsim \pi^2/4$, with $\rho$ its average density. Equivalently, the compactness of the composite body should satisfy
${\cal C}_{\rm global}\gtrsim {\cal C}_{\rm crit}$~\footnote{Note that the compactness of a composite body can be made larger than that of any of its individual constituent bodies.}.
This is the main purpose of this study, to test and challenge this result.
For example, suppose we take a composite body which is a cube of side ${\cal L}$ made of $N$ spheres (each of radius $R$ and density $\rho_0$) equally distributed within the cube. Then, the scalarization condition~\eqref{eq_instability_sphere} can be used, with the replacement $\mu^2 \to -4\pi\alpha N\rho_0 R^3/(3{\cal L}^3)$.
For the equivalent total size we can either use a circumscribed or inscribed sphere, yielding ${\cal L}/\sqrt{2}$ or ${\cal L}/2$ respectively. One finds the condition
\be
\frac{NR^3}{{\cal L}}\gtrsim -\frac{3\pi}{(6\pm 2)\alpha\rho_0}=\frac{3\pi}{(6\pm 2)\mu^2}\gtrsim\frac{1.2}{\mu^2}\,. \label{scalar_infinite}
\ee
Can this condition be violated both for a single sphere {\it and} for the entire lattice?
To investigate this issue we have investigated three models of matter.

%%%%%%%%%%%%%%%%%%%%%%%%%%%%%%%%%%%%%%%%%%%%%%%%%%%%%%%%%%%%%%%%%
%\subsection{The concentric shell model}
%%%%%%%%%%%%%%%%%%%%%%%%%%%%%%%%%%%%%%%%%%%%%%%%%%%%%%%%%%%%%%%%%
%%%%%%%%%%%%%%%%%%%%%%%%%%%%%%%%%%%%%%%%%%%%%%%%%%%%%%%%%%%%%%%%%%%%%
\noindent{\bf{\em I. The concentric shell model.}}
%%%%%%%%%%%%%%%%%%%%%%%%%%%%%%%%%%%%%%%%%%%%%%%%%%%%%%%%%%%%%%%%%%%%%
Analytic results describing realistic bodies are hard to devise. One can make progress by 
building a spherically-symmetric arrangement with radial structure.
Consider a sphere of radius $R$ and density $-\alpha\rho_0=\mu^2$, surrounded by a series of $N$ concentric infinitely-thin shells of radius $r_i=R+i2\gamma \beta R$, and density $\rho^{\delta}_i=\gamma R\mu^2\delta(r-r_i)$. This model can be worked out analytically when $\gamma\ll 1$. We start with a single sphere at the critical radius given by Eq.~\eqref{eq_instability_sphere}. At this radius, an equilibrium solution of nonzero scalar inside the sphere is possible. We look for deformations induced by the outer shells, in a marginal stability situation. We find that to order ${\cal O}(\gamma^2)$, 
\beq
\mu^2 R^2&=&\frac{\pi^2}{4}\left(1-2N\gamma\right)\,,\\
\rho_j r_j^2&=&\frac{\pi^2}{4}\left(1-\gamma(2j\beta+2N-3j)\right)\,,
\eeq
where $\rho_j$ is the average density of the configuration including all shells up to $j\leq N$. 

In other words, this is an example of a composite system where the critical threshold for scalarization can be lowered
by increasing the number of shells $N$ (and as long as $\beta>1/2$). Notice that the compactness of the central sphere as well as of each shell $\rho_j r_j^2$ decrease when $N$ increases.
The decrease in compactness is not arbitrary, since terms of order $N\gamma$ were assumed to be small in the calculation above.
Numerical results indicate that there are, in fact, lower limits on the compactness of the configuration for it to be scalarized.

%%%%%%%%%%%%%%%%%%%%%%%%%%%%%%%%%%%%%%%%%%%%%%%%%%%%%%%%%%%%%%%%%
%\subsection{The Wigner-Seitz model for a periodic lattice}
%%%%%%%%%%%%%%%%%%%%%%%%%%%%%%%%%%%%%%%%%%%%%%%%%%%%%%%%%%%%%%%%%
%%%%%%%%%%%%%%%%%%%%%%%%%%%%%%%%%%%%%%%%%%%%%%%%%%%%%%%%%%%%%%%%%%%%%
\noindent{\bf{\em II. The Wigner-Seitz model for a periodic lattice.}}
%%%%%%%%%%%%%%%%%%%%%%%%%%%%%%%%%%%%%%%%%%%%%%%%%%%%%%%%%%%%%%%%%%%%%
To study realistic composite bodies, the assumption of spherical symmetry must be discarded.
Consider now a lattice of small objects, describing for instance atoms or nuclei.
When the lattice is infinite, the problem can be reduced to a simpler one of a
single object with appropriate boundary conditions at the midpoint with its nearest neighbours.
The classical work by Wigner and Seitz (hereafter ``WS'') showed that simple Neumann conditions on an appropriate cell
describe the full problem, and how in certain limits that cell can be approximated by a sphere centered on the object~\cite{Wigner:1933zz}. Our problem is formally equivalent to theirs, and we consider the same, WS approximation.

Take then a periodic lattice of spheres, each of radius $R$ (and density $\mu^2=-\alpha \rho_0$) separated by a distance $2L$.
The problem can be formally solved by imposing regularity at the center of one sphere and Neumann conditions on the scalar at a distance
$L$. One finds the eigenvalue problem for the eigenfrequency $\omega$
\beq
&&e^{2iR\omega}\omega L_{-}-e^{2iL\omega}\omega L_{+}+\sqrt{\mu^2+\omega^2}\cot{(R\sqrt{\mu^2+\omega^2})}\nonumber\\
&\times&i\left(e^{2iL\omega}L_{+}+e^{2iR\omega}L_{-}\right)=0\,.
\eeq
Here, $L_\pm=(L\omega\pm i)$. We find the following description of the numerical roots of the above equation,
\beq
\omega&\approx& i\mu \left(\frac{R}{L}\right)^{3/2}\,,\qquad R\mu\ll \pi/2\,,\label{eq:Wigner}\\
\omega&\approx&iL^{-1} \,,\qquad  \qquad R\mu \sim \pi/2\,. 
\eeq
This result indicates that an infinite array of {\it dilute} bodies (where condition \eqref{eq_instability_sphere} is violated individually for each element of the lattice) is always unstable,
suggesting scalarization.

Of course, Eq.~\eqref{scalar_infinite} is always satisfied for a periodic but spatially infinite distribution. Thus, even though each individual sphere is not scalarized, the entire setup is just because it is compact enough. 

The WS construction can be extrapolated to finite-sized structures as long as the wavelength of the unstable mode is shorter than the total sample size, i.e., as long as $2\pi/|\omega|\lesssim{\cal L}$ which can be re-written as,
\beq
\frac{NR^3}{{\cal L}}&\gtrsim&  \frac{\pi^2}{2\mu^2}\,,\qquad R\mu\ll \pi/2\,,\label{eq:wigner1}\\
L&\lesssim& {\cal L} \,,\qquad R\mu\sim \pi/2\,.
\eeq
The second condition is trivially satisfied. Thus, when the individual spheres are close to the critical value for scalarization,
collective effects always work to scalarize composite bodies, even if their average compactness is below the threshold. 
On the other hand, Eq.~\eqref{eq:wigner1} is suggestively close to the original bound~\eqref{scalar_infinite}. This indicates that when the building blocks (the individual spheres) are far from the threshold, then the only way to scalarize a composite body is for it to satisfy, on the average, a scalarization condition. However, one needs to turn to fully numerical calculations for conclusive results.

\begin{table}[h]
 \caption{Summary of our results for the instability threshold when $N$ equally spaced spheres are arranged in a cube. We vary the cube lateral length ${\cal L}$ and determine the threshold ${\cal L}_{\rm crit}$ above which the configuration is stable. We add also the corresponding (minimum) critical distance between spheres, which carry errors of order $\sim 2\%$ or less. Our units are such that $\mu^2=1$. The critical compactness $NR^3/{\cal L}_{\rm crit}$ violates threshold Eq.~\eqref{scalar_infinite} for $R\mu \gtrsim 1$, but asymptotes to it for small $R\mu$.\label{tab:Lcrit}}
  \begin{ruledtabular}
    \begin{tabular}{ccccc}
      $N$  &  $R$  &  $d_{\rm crit}$     & ${\cal L}_{\rm crit}$      & $NR^3/{\cal L}_{\rm crit}$\\
      \hline
      8    &  1.0  &  $1.3$             & $3.3  \pm 0.1$      & $2.4\pm 0.1$\\
      27   &  1.0  &  $6.6$             & $17.25\pm 0.25$     & $1.56\pm 0.02$\\
      64   &  1.0  &  $24.8$            & $80.5 \pm 0.5$      & $0.80\pm 0.01$\\
      125  &  1.0  &  $>48$             & $>200$              & $<0.62$\\ \hline
%
%      8    &  0.5 &  & $<1.4$              & $>0.7$\\
%      27   &  0.5 &  & $<2.5$              & $>1.35$\\
      64   &  0.5  &  $0.5 $             & $4.4 \pm 0.1$       & $1.81\pm 0.05$\\
      125  &  0.5  &  $1.4$              & $9.4 \pm 0.2$       & $1.66\pm0.04$\\
      216  &  0.5  &  $2.4$              & $17 \pm 0.2$        & $1.58\pm0.03$\\
      343  &  0.5  &  $3.7$              & $28 \pm 0.3$        & $1.53\pm  0.02$\\
      729  &  0.5  &  $7.5$              & $59.2 \pm 0.2$      & $1.53\pm  0.01$\\
    \end{tabular}
  \end{ruledtabular}
\end{table}
\begin{figure*}[ht!]
  \centering
  \includegraphics[width=0.32\textwidth]{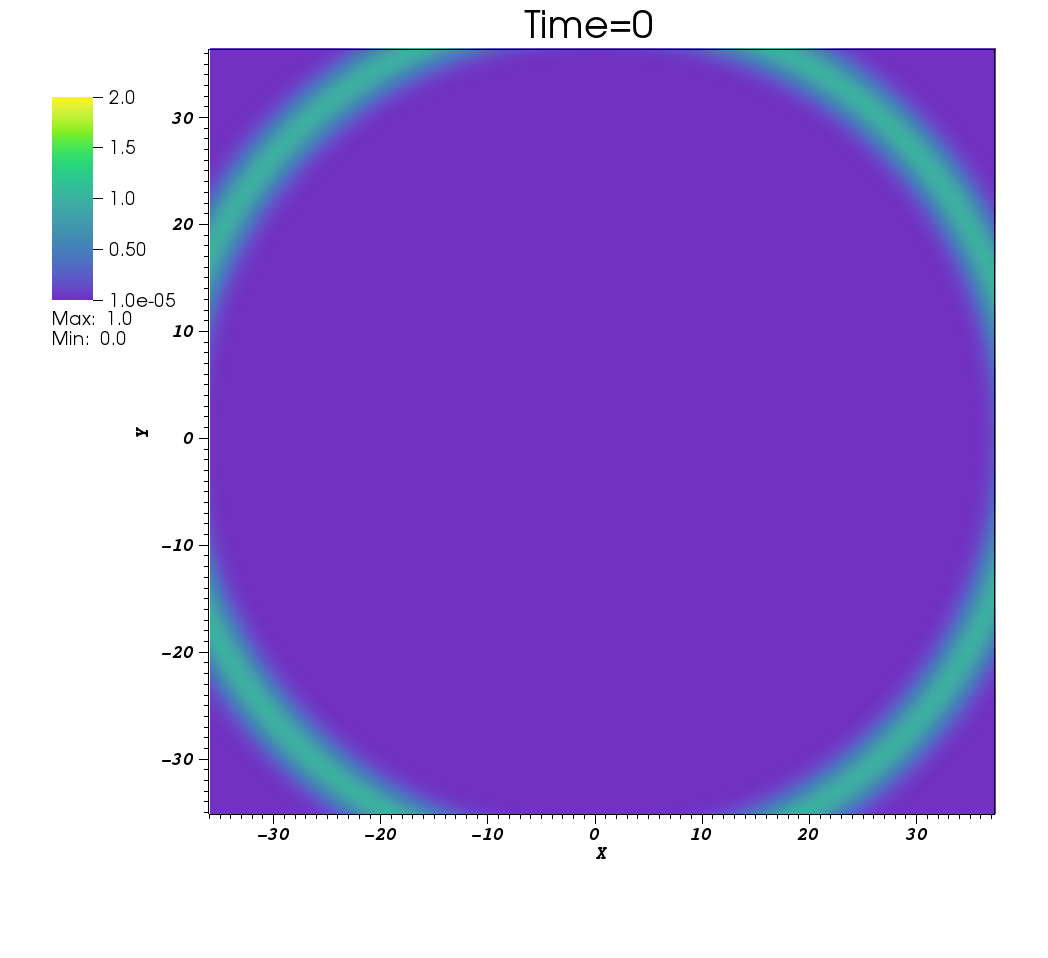}
  \includegraphics[width=0.32\textwidth]{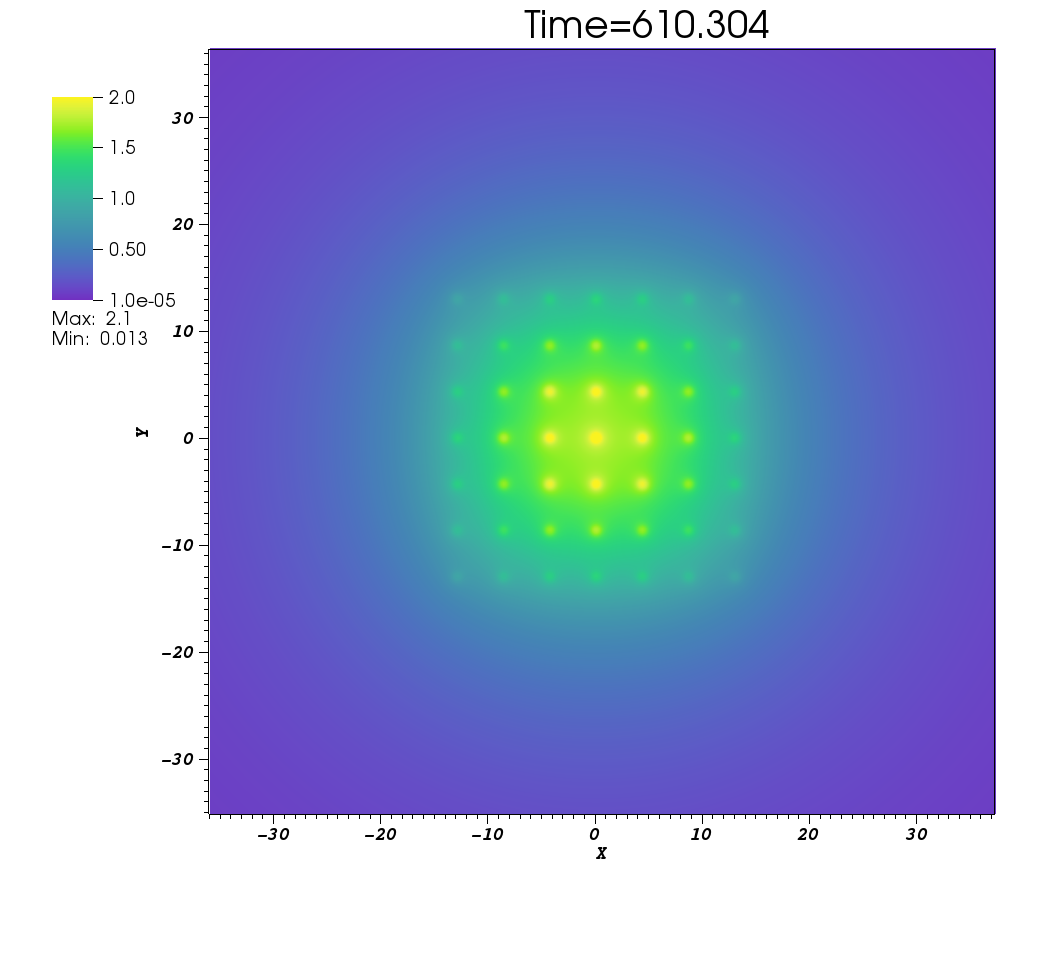}
  \includegraphics[width=0.32\textwidth]{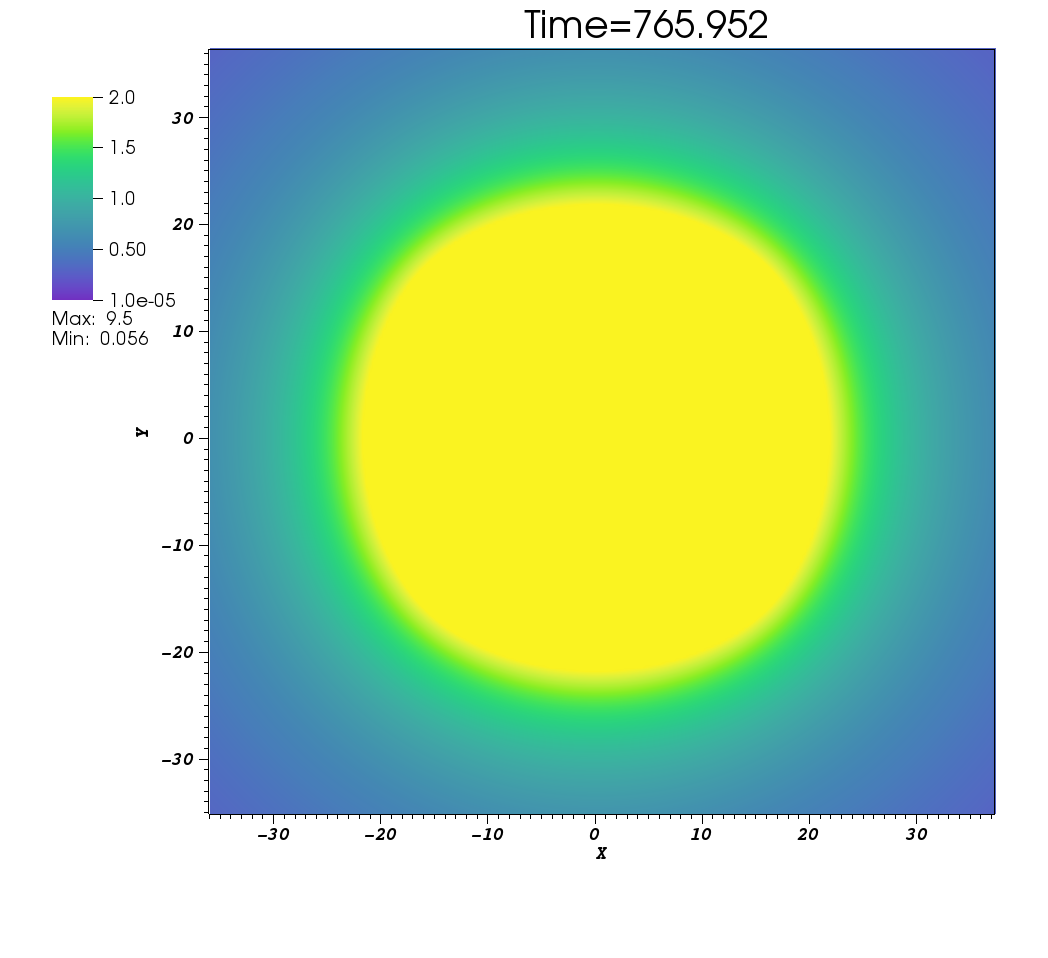}
\caption{Snapshots of the evolution of a gaussian wavepacket around a cube of lateral size ${\cal L}\mu=26$ composed of $7^3=343$ spheres, each of radius $R\mu=0.5$. Color intensity depicts magnitude of the scalar field. The left panel shows the initial data. After $t\mu \sim 610$ (middle panel), the scalar is localized at the spheres and growing. After $t\mu \sim 766$ the scalar is now ``seeing'' the entire cubic cluster and still growing exponentially. 
\label{Fig:N7_R05}}
\end{figure*}
\begin{figure}[ht!]
  \centering
  \includegraphics[width=0.5\textwidth]{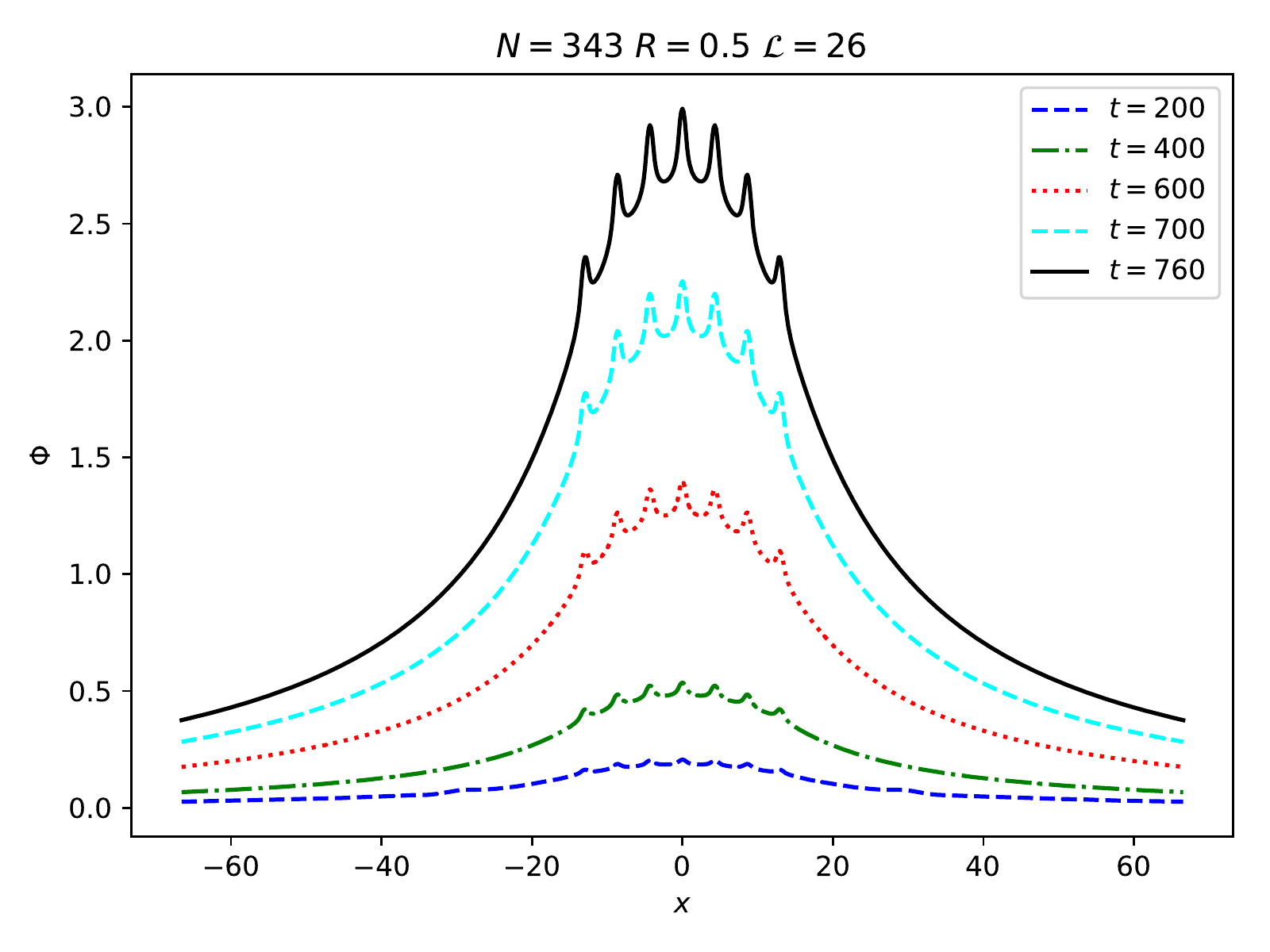}
\caption{Scalar field profile of an unstable lattice, with 343 spheres (arranged in layers of 7), each of radius $R\mu=0.5$ arranged in a cube of side ${\cal L}\mu=26$. The field is measured on the $y=z=0$ plane. Notice that the field derivative vanishes at the points between spheres. \label{Fig:N7_R05_profile}}
\end{figure}
%
%%%%%%%%%%%%%%%%%%%%%%%%%%%%%%%%%%%%%%%%%%%%%%%%%%%%%%%%%%%%%%%%% 
%\subsection{A finite cubic lattice: numerical results}
%%%%%%%%%%%%%%%%%%%%%%%%%%%%%%%%%%%%%%%%%%%%%%%%%%%%%%%%%%%%%%%%%
%%%%%%%%%%%%%%%%%%%%%%%%%%%%%%%%%%%%%%%%%%%%%%%%%%%%%%%%%%%%%%%%%%%%%
\noindent{\bf{\em III. Numerical results for a cubic lattice.}}
%%%%%%%%%%%%%%%%%%%%%%%%%%%%%%%%%%%%%%%%%%%%%%%%%%%%%%%%%%%%%%%%%%%%%
Consider therefore an array of $N$ equally spaced spheres, each of radius $R$ arranged in a cubic fashion.
We model such a system through the equation
\beq
\nabla_\alpha\nabla^\alpha \Phi(t,\vec r) &=& -\mu^{2}f_{\mu}(\vec r) \Phi(t,\vec r) \,, \label{eq:KG} \\
f_{\mu}(\vec r) & \equiv& \sum_{a=1}^{N} \mathcal{S}(|\vec r - \vec r_a|)\,,\label{eq:fmu}
\eeq
where $\vec r_a$ % \equiv (x_a, y_a, z_a)$, $x_a = -\frac{L}{2} + (a-1) \frac{L}{N-1}$, $a=1,\ldots,N$
is the center of each sphere and the mass-function $\mathcal{S}$ should be unity inside the spheres and vanishingly small away from them. Numerically, we approximate this function by
\be
\mathcal{S}(r) \equiv \frac{1}{2} \left[ \tanh{\left(h_s(r+R)\right)}
  - \tanh{\left(h_s(r-R)\right)} \right] \,,
\ee
where we typically fix $h_s = 10$.

To perform numerical evolutions of this system we make use of the
\textsc{EinsteinToolkit}
infrastructure~\cite{Loffler:2011ay,maria_babiuc_hamilton_2019_3522086,Zilhao:2013hia},
with mesh-refinement capabilities handled by the \textsc{Carpet}
package~\cite{Schnetter:2003rb,CarpetCode:web}. Typically we use four
refinement levels, with resolutions spanning $\Delta x = 0.16$ in the region
where the spheres are located to $\Delta x = 1.28$ at the outer boundary.
We use a cubic grid with size $L_x = L_y = L_z = 128$. At the outer boundary we impose Sommerfeld boundary conditions.
Due to
memory constraints, it becomes increasingly challenging to probe configurations
with many spheres, where large grids are required.
We initialize the system with a simple Gaussian pulse,
\[
  \Phi(0,r) = e^{-\frac{(r-r_0)^2}{2\sigma^2}} ,
\]
where we typically fix $\sigma = 8$ and $r_0 = 40$, as seen in the first panel of Fig.~\ref{Fig:N7_R05}.

Our results show convergence between second and third order. We have further tested our setup and numerical implementation by evolving a single sphere, and fixing $\mu^2=1$.
For large sphere radius we find that the scalar increases exponentially with time everywhere, whereas for small sphere radii the initial fluctuation dies off. Our results indicate that the critical sphere radius for $N=1$ configurations is $R=1.58 \pm 0.02$, in agreement with the analytical prediction \eqref{eq_instability_sphere}.

Table~\ref{tab:Lcrit} summarizes our results for the evolution of scalar fields around cubic-like arrangements of spheres. 
We have tested the predictions from the WS approximation, and considered two regimes, a near-critical with $R\mu=1$ and a regime where each sphere is far from the threshold, with $R\mu=0.5$. Our results are fully consistent with the simple WS model:
composite configurations of $R\mu=1$ spheres can easily scalarize, the entire lattice having a compactness which is far from the threshold for scalarization. An averaging of the lattice properties, Eq.~\eqref{scalar_infinite}, would predict no scalarization where, in fact, there is one.

Focus now on arrangements where each sphere is far from the scalarization threshold, $R\mu=0.5$. This is, potentially the interesting regime for new phenomenology. However, our results indicate -- in line with the WS approximation -- that to scalarize the entire arrangement then its compactness must satisfy the constraint~\eqref{scalar_infinite}. Snapshots of a marginally unstable configuration are shown in Fig.~\ref{Fig:N7_R05}. The initial ingoing pulse is seen to produce scalar excitations, initially localized at each sphere and which then grow exponentially as a whole, i.e., supported on the entire lattice.
The reason for the accuracy of the WS simple model is supported in Fig.~\ref{Fig:N7_R05_profile}, where we show the scalar profile in the $x$-axis. Neumann conditions are indeed a good description of the field in the midpoints, even for such a relatively low number of spheres (layers of 7 spheres, a total of 343 spheres).

%%%%%%%%%%%%%%%%%%%%%%%%%%%%%%%%%%%%%%
\noindent{\bf{\em Conclusions.}}
%%%%%%%%%%%%%%%%%%%%%%%%%%%%%%%%%%%%%%
Our results are clear. When a body of constant density is close to, but below the scalarization threshold~\eqref{eq_instability_sphere}
it does not scalarize {\it individually}. However, a collection of such bodies can easily scalarize, while maintaining the average
compactness much below the threshold~\eqref{scalar_infinite}. 
A intuitive reason, born out of our results, seems to be that individually-stable objects are still able to excite modes whose Compton wavelength
is large enough to interfere with its neighbours. When these are close enough to catch the tail of such modes then collective scalarization can occur. Such crude argument requires a deeper understanding to be made quantitative.
However, collective behavior does not have a significant impact on our ability to constraint scalar-tensor theories in the sense that a composite body obeys, to a good extent~\eqref{scalar_infinite} in an average sense, when its fundamental building blocks are far from it. In other words, our results show that it is legitimate -- as far as realistic situations go -- to take the local average of matter properties to investigate scalarization.

%%%%%%%%%%%%%%%%%%%%%%%%%%%%%%%%%%%%%%%%%%%%%%%%%%%%%%%%%%%%%%%%%%%%%%%%%%%%%
\section*{Acknowledgements}
%%%%%%%%%%%%%%%%%%%%%%%%%%%%%%%%%%%%%%%%%%%%%%%%%%%%%%%%%%%%%%%%%%%%%%%%%%%%%
%
We are thankful to J. L. Martins for useful discussions and for directing us to relevant solid-state physics
references. We are thankful to an anonymous referee for many and useful suggestions.
We thank Lorenzo Annulli and Taishi Ikeda for helpful discussions and for sharing some of their numerical results.
We are indebted to Waseda University for warm hospitality while this work was being finalized.
V.~C.\ acknowledges financial support provided under the European Union's H2020 ERC 
Consolidator Grant ``Matter and strong-field gravity: New frontiers in Einstein's 
theory'' grant agreement no. MaGRaTh--646597.
M.~Z.\ acknowledges financial support provided by FCT/Portugal through the IF programme, grant IF/00729/2015.
This project has received funding from the European Union's Horizon 2020 research and innovation 
programme under the Marie Sklodowska-Curie grant agreement No 690904.
We thank FCT for financial support through Project~No.~UIDB/00099/2020.
We acknowledge financial support provided by FCT/Portugal through grant PTDC/MAT-APL/30043/2017.
The authors would like to acknowledge networking support by the GWverse COST Action 
CA16104, ``Black holes, gravitational waves and fundamental physics.''
%
%%%%%%%%%%%%%%%%%%%%%%%%%%%%%%%%%%%%%%%%%%%%%%%%%%%%%%%%%%%%%%%%%%%%%%%%%%%%%

% \bibliographystyle{apsrev4}
% \bibliographystyle{utphys}
\bibliography{References}
\end{document}